\documentclass[reprint,prl,aps,nofootinbib,nobalancelastpage]{revtex4-2}
\usepackage{amsmath}
\usepackage{amsfonts}
\usepackage{amssymb}
\usepackage{amsthm}
\usepackage{mathtools}
\usepackage{graphicx}
\usepackage[dvipsnames]{xcolor}
\usepackage[colorlinks=True,citecolor=myBlue,linkcolor=myRed,urlcolor=myBlue,hypertexnames=false]{hyperref}
\usepackage{hypcap}
\usepackage{yfonts}
\usepackage{bm}
\usepackage[normalem]{ulem}
\usepackage{dsfont}
\usepackage{braket}
\usepackage{float}
\usepackage{marginnote}
\usepackage{tikz}
\usepackage{quantikz}
\usetikzlibrary{shapes, arrows.meta,decorations.pathreplacing}

\newcommand{\mbf}[1]{\mathbf{#1}}

\usepackage{kantlipsum}

\newcommand\eq[1]{\begin{align}#1\end{align}}

\definecolor{myBlue}{RGB}{31,119,180}
\definecolor{myOrange}{RGB}{255,127,14}
\definecolor{myGreen}{RGB}{44,160,44}
\definecolor{myRed}{RGB}{214,39,40}
\definecolor{myPurple}{RGB}{148,103,189}

\newcommand\mytitle{
Quantum Chaos and Diffusive Transport from Geometric Randomness}

\begin{document}

\title{\mytitle}

\author{Bibek Saha}
\email{bibek.saha@icts.res.in}
\affiliation{International Centre for Theoretical Sciences, Tata Institute of Fundamental Research, Bengaluru 560089, India}

\author{Abhishek Dhar}
\email{abhishek.dhar@icts.res.in}
\affiliation{International Centre for Theoretical Sciences, Tata Institute of Fundamental Research, Bengaluru 560089, India}

\author{Sthitadhi Roy}
\email{sthitadhi.roy@icts.res.in}
\affiliation{International Centre for Theoretical Sciences, Tata Institute of Fundamental Research, Bengaluru 560089, India}

\begin{abstract}
The physics of quantum chaos and diffusive transport is typically studied in settings with microscopic disorder or many-body interactions. In this Letter, we demonstrate that these phenomena can arise purely from geometric randomness. By studying non-interacting quantum particles on random locally tree-like layered graphs with uniform couplings, we show that the geometric randomness and effective graph dimensionality dictates the presence of chaotic dynamics or lack thereof. These graphs can be considered as structurally disordered generalisations of regular square lattices or ladders, or equivalently as multi-component one-dimensional chains with random links between the components. We find that an extensive layer size yields robust quantum chaos, level repulsion, and diffusive transport. Conversely, in the quasi-one-dimensional limit, we find the coexistence of extensive number of localised and delocalised states -- this leads to suppressed level repulsion accompanied by the latter driving ballistic transport. These results establish geometric randomness as a fundamental and independent mechanism for generating and tuning quantum chaos.
\end{abstract}

\maketitle

\paragraph*{Introduction:} 
The emergence of quantum chaos or lack thereof, manifested in localisation, has been for long and continues to be one of the cornerstones of condensed matter and statistical physics~\cite{stockmann1999quantum,haake2010quantum}. 
While this question has received a lot of attention in the recent past in the context of interacting, disordered quantum many-body systems under the umbrella of eigenstate thermalisation and many-body localisation~\cite{rigol2008thermalisation,nandkishore2015many,dalessio2016from,deutsch2018eigenstate,abanin2019colloquium,sierant2025many}, the fundamental problem of non-interacting quantum particles in random geometries continues to be the beating heart of this question. 
The reason for this is broadly twofold. 

First, non-interacting particles on high-dimensional, disordered graphs are fertile grounds for studying quantum chaos and transitions to localisation through Anderson transitions with exotic behaviour such as multifractality and anomalous transport appearing en route~\cite{abou-chacra1973self,tikhonov2016anderson,Tikhonov2016CayleyTree,garciamata2017scaling,detomasi2019survival,detomasi2020subdiffusion,roy2020localisation,biroli2020anomalous,garciamata2022critical}. 
Second, the problem of many-body chaos and many-body localisation in interacting systems can be mapped onto a problem of a fictitious single particle on the high-dimensional Fock-space graph of the interacting system~\cite{logan1990quantum,altshuler1997quasiparticle,biroli2017delocalised,logan2019many,tarzia2020manybody,roy2020fock,detomasi2020rare,roy2021fockspace,tikhonov2021eigenstate,tikhonov2021from,roy2024fock}.
In all of these studies, the key ingredients are the high-dimensional nature of the graphs and  the presence of disorder in the microscopic couplings of the Hamiltonian, the strength of which in fact tunes the model from chaotic to localised. 

While chaos and transport are intricately connected, chaos does not 
necessarily imply diffusive transport; a well-known counterexample being the classical Fermi-Pasta-Ulam-Tsingou chain~\cite{lepri1997}.  In quantum systems, chaos and diffusive transport has been demonstrated in interacting, weakly disordered spin- and fermionic chains~\cite{agarwal2015anomalous,luitz2016anomalous,khait2016spin,znidaric2016diffusive}, in a chain of coupled SYK dots~\cite{gu2017}, and more recently in a random matrix model with spatial extent~\cite{chalker2025}. 
Diffusion has also been demonstrated in a non-interacting linear chain with an $N$-component wavefunction with random matrices connecting the components on neighbouring sites~\cite{mudry2000}. 
However, the matter of chaos and diffusion arising purely from geometric randomness has remained hitherto unexplored.

In this Letter, we address the question: can models on finite-dimensional graphs with non-random microscopic couplings show the rich array of behaviour from chaos to localisation purely due to geometric randomness in the graph?
Our analysis is based on ``random locally tree-like layered'' (RLTL) graphs~\cite{dhar2011hard}. 
They constitute a family of graphs that possess a regular spatial structure as the sites are organised along layers.
At the same time, the links between adjacent layers are random which endows these graphs with geometric randomness -- this also leads to the graphs being locally tree like.

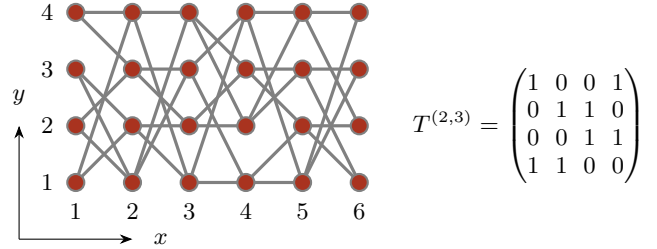
\begin{figure}
\begin{tikzpicture}[scale=0.75]
\draw[gray,very thick] (1,1) -- (2,2);
\draw[gray,very thick] (1,1) -- (2,4);
\draw[gray,very thick] (1,2) -- (2,3);
\draw[gray,very thick] (1,2) -- (2,1);
\draw[gray,very thick] (1,3) -- (2,2);
\draw[gray,very thick] (1,3) -- (2,1);
\draw[gray,very thick] (1,4) -- (2,3);
\draw[gray,very thick] (1,4) -- (2,4);
\draw[gray,very thick] (2,1) -- (3,3);
\draw[gray,very thick] (2,1) -- (3,4);
\draw[gray,very thick] (2,2) -- (3,2);
\draw[gray,very thick] (2,2) -- (3,1);
\draw[gray,very thick] (2,3) -- (3,2);
\draw[gray,very thick] (2,3) -- (3,3);
\draw[gray,very thick] (2,4) -- (3,4);
\draw[gray,very thick] (2,4) -- (3,1);
\draw[gray,very thick] (3,1) -- (4,4);
\draw[gray,very thick] (3,1) -- (4,1);
\draw[gray,very thick] (3,2) -- (4,3);
\draw[gray,very thick] (3,2) -- (4,2);
\draw[gray,very thick] (3,3) -- (4,4);
\draw[gray,very thick] (3,3) -- (4,1);
\draw[gray,very thick] (3,4) -- (4,3);
\draw[gray,very thick] (3,4) -- (4,2);
\draw[gray,very thick] (4,1) -- (5,2);
\draw[gray,very thick] (4,1) -- (5,1);
\draw[gray,very thick] (4,2) -- (5,4);
\draw[gray,very thick] (4,2) -- (5,3);
\draw[gray,very thick] (4,3) -- (5,2);
\draw[gray,very thick] (4,3) -- (5,3);
\draw[gray,very thick] (4,4) -- (5,4);
\draw[gray,very thick] (4,4) -- (5,1);
\draw[gray,very thick] (5,1) -- (6,3);
\draw[gray,very thick] (5,1) -- (6,4);
\draw[gray,very thick] (5,2) -- (6,1);
\draw[gray,very thick] (5,2) -- (6,2);
\draw[gray,very thick] (5,3) -- (6,1);
\draw[gray,very thick] (5,3) -- (6,3);
\draw[gray,very thick] (5,4) -- (6,4);
\draw[gray,very thick] (5,4) -- (6,2);

\foreach \x in {1,3,5}{
    \node at (\x,0.5) {$\x$};
    \foreach \y in {1,...,4}{
    \filldraw[draw=gray,thick, fill=Mahogany] (\x,\y) circle (0.15);
    }
}
\foreach \x in {2,4,6}{
    \node at (\x,0.5) {$\x$};
    \foreach \y in {1,...,4}{
    \filldraw[draw=gray,thick, fill=Mahogany] (\x,\y) circle (0.15);
    }
}
\foreach \x in {1,...,4}{
    \node at (0.5,\x) {$\x$};
}
\draw[-Stealth] (0,0) -- (2,0);
\draw[-Stealth] (0,0) -- (0,2);
\node at (2.5,0) {$x$};
\node at (0,2.5) {$y$};

\node at (9,2) {$T^{(2,3)} =\begin{pmatrix}
1 & 0 & 0 & 1\\
0 & 1 & 1 & 0\\
0 & 0 & 1 & 1\\
1 & 1 & 0 & 0
\end{pmatrix}$};

\end{tikzpicture}
\caption{{\bf RLTL graph}: An instance of the graph with $L_x=6$ and $L_y=4$, and $K=2$. For illustration, the hopping matrix between layers $x=2$ and $3$ has also been specified explicitly. }
\label{fig:rltl}
\end{figure}

\paragraph*{Main results:} We find that the effective dimensionality of the graph drastically alters the nature of the eigenstates and dynamics on the graph.
Specifically, we consider two cases. In the first case, the ratio of the number of sites in a given layer and the number of layers remains finite in the thermodynamic limit -- in this case, we find the emergence of chaos throughout the spectrum (except at energy $E=0$) manifested in level repulsion accompanied by delocalised states and diffusive transport. In this regime, we also find a diffusive growth of bipartite entanglement.
On the other hand, in the second case, we consider the number of sites in a given layer to be an $O(1)$ constant, and take the thermodynamic limit by taking the number of layers to infinity -- in this case the spectrum has an extensive number of both localised and delocalised states with the latter occurring at fixed, isolated energies independent of the RLTL graph realisation. 
The presence of the localised states suppresses the level repulsion in the spectrum.
At the same time, the extensive number of delocalised states does lead to ballistic transport whose spatial profile is anomalous. 
These results therefore show how quantum chaos and diffusive transport can emerge purely due to geometric randomness and the influence of the effective dimensionality of the graph on them.

The remainder of the paper is organised as follows.
We start with describing concretely the model of RLTL graphs.
We then present results for the density of states and level spacing statistics to diagnose chaos, or lack thereof, in the spectrum.
This is accompanied by results on inverse participation ratios (IPRs) of eigenstates to probe their (de)localised nature. 
To show the presence of anomalous transport, we then present results on wavepacket dynamics.
Finally, we also present results on bipartite entanglement dynamics to supplement the transport results.

\paragraph*{Model:} 
The family of RLTL graphs with connectivity $2K$ is described by organising the sites along layers, labelled by $x$, such that any site on layer $x$ is randomly connected to $K$ distinct sites on layer $x-1$ and similarly on $x+1$.
The tight-binding Hamiltonian describing non-interacting fermions on the graph is given by
\eq{
H = J\sum_{x=1}^{L_x-1} \hat{\Psi}^\dagger_{x+1} T^{(x,x+1)} \hat{\Psi}_{x} + {\rm h.c.}\,, 
\label{eq:ham}
}
with $\hat{\Psi}_{x} = [c_{x,1},c_{x,2},\cdots,c_{x,L_y}]^{\mathsf{T}}$ where $c_{x,y}$ is the annihilation operator on site $y$ of layer $x$ and $J$ is the hopping amplitude which we set to 1.
The set of $L_y\times L_y$ matrices, $\{T^{(x,x+1)}\}$, where each row/column in any such matrix has exactly $K$ entries which are 1 and the rest are 0, specifies an instance of the RLTL graph. 
An example is shown in Fig.~\ref{fig:rltl}. Our model can also be interpreted as a chain of $L_x$ sites, with a $L_y$-component wavefunction at each site and random connections between the components on neighbouring sites.
We will consider two cases: (i) the ratio $W=L_y/L_x$ stays finite as the thermodynamic limit is taken which corresponds effectively to a RLTL graph in two dimensions, and (ii) $L_y\sim O(1)$ constant while the thermodynamic limit is taken by $L_x\to\infty$, which corresponds to an effective one-dimensional RLTL ladder with $L_y$ legs.
As representatives of the two cases, we will consider $W=1/2$ and $L_y=8$, and set $K=2$ throughout the rest of the paper.

\begin{figure}
\includegraphics[width=\linewidth]{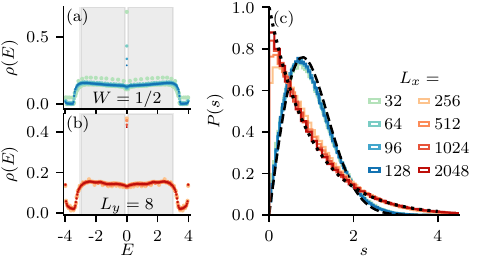} 
\caption{{\bf Spectrum and level statistics:} (a) and (b) show the normalised DoS for $W=1/2$ and $L_y=8$ respectively. In both cases, the DoS is approximately flat and the peak at $E=0$ decays with increasing system size. (c) The distribution of the normalised level spacings (defined in Eq.~\eqref{eq:level-spacing}). Blue and red tones correspond to $W=1/2$ and $L_y=8$ respectively with darker colours denoting larger $L_x$. The black dashed and dotted lines denote the Wigner surmise and Poisson distribution respectively.}
\label{fig:spectrum}
\end{figure}

Note that the RLTL graph is bipartite as all sites on the odd layers can be considered to be of a particular sublattice and all sites on the even layers to be of the other.
Given that the hoppings in the model in Eq.~\eqref{eq:ham} are only between two different sublattices, and there are no onsite potentials, the model manifestly has a sublattice symmetry.
This endows the spectrum with a $E\to -E$ symmetry; for every eigenenergy $E$ there must exist another at $-E$, and the two eigenstates are related to each other via a sublattice transformation.
In such a situation, one may expect a Dyson-like singularity~\cite{dyson1953dynamics,fleishman1977fluctuations,krapivsky2011dynamics,detomasi2016generalised} in the density of states (DoS) at $E=0$. 
However, we find that the peak in the DoS at $E=0$, while present for finite systems, decays in height with increasing system size (see Fig.~\ref{fig:spectrum}), indicating the absence of such a singularity in RLTL graphs.

\paragraph*{Spectrum:}

To probe the presence of chaos in the spectrum, we study the level-spacing statistics, $P(s)$, where $s$ is defined as 
\eq{
s_i = (E_{i+1}-E_i)\times {\rho\left(\overline{E}_{i,i+1}\right)}\,.
\label{eq:level-spacing}
}
Here $\rho(E)$ is the normalised DoS at energy $E$
and $\overline{E}_{i,i+1} = (E_i+E_{i+1})/2$. 
Note that multiplying the successive energy differences by the DoS, instead of unfolding the spectrum, is sufficient in this case as the DoS is flat to an excellent approximation.
The results for $P(s)$ are shown in Fig.~\ref{fig:spectrum}.
For $W=1/2$, the data are extremely well described by the Wigner surmise~\cite{mehta2014random}, $P(s) = \pi s e^{-s^2/4}/2$.
This indicates level repulsion of the random matrix kind, signifying the presence of quantum chaos.
By contrast, for a fixed $L_y=8$, the data for $P(s)$ drifts towards the Poissonian behaviour of $P(s) = e^{-s}$ with increasing $L_x$ indicating the absence of level repulsion and chaos in the spectrum, in the thermodynamic limit. 
This is the first evidence of the presence of chaos in the effective two-dimensional graph and lack thereof in the effective ladder geometry.

\begin{figure}
\includegraphics[width=\linewidth]{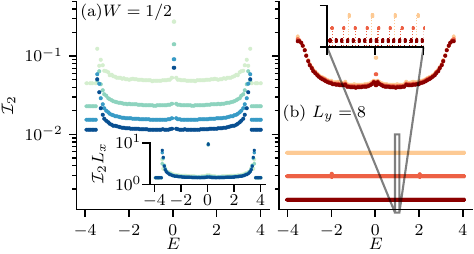} 
\caption{{\bf IPRs as a function of $\bm{E}$:} (a) For $W=1/2$, the IPR scales as $L_x^{-1}$ at all $E$ as evinced by the collapse of the data for ${\cal I}_2L_x$ (inset) for different $L_x=32,64,96,128$ (lighter to darker colours). (b) For $L_y=8$, the unscaled IPR collapses for different $L_x=256,512,1024$ (lighter to darker) indicating ${\cal I}_2\sim L_x^0$ and hence localisation, except for the Bloch-wave like states. The inset shows an enlargement of the small energy window where the vertical dashed lines denote the Bloch-state energies.} 
\label{fig:ipr}
\end{figure}

\paragraph*{Eigenstates:}
We now turn towards the eigenstates, in particular, their (de)localisation properties quantified by the IPRs. 
Since the RLTL graphs have a notion of locality only across layers (along the $x$ direction in Fig.~\ref{fig:rltl}), we define the IPR also across the layers as
\eq{
{\cal I}_2 = \sum_{x=1}^{L_x}\phi^2(x)\,;~~\phi(x) = \sum_{y=1}^{L_y}|\psi(x,y)|^2\,,
}
where $\psi(x,y) = \braket{x,y|\psi}$ is the amplitude of the eigenstate $\ket{\psi}$ at the site $(x,y)$.
The results for the IPR are shown in Fig.~\ref{fig:ipr}.

For finite $W$, the data show that the IPR scales as ${\cal I}_2\sim L_x^{-1}$ throughout the spectrum, signifying delocalised states concomitant with level repulsion. 
The exception to this is at $E=0$ where the states show (quasi)localised behaviour; however, we find that these states are localised near the boundaries of the RLTL graph at $x=1$ and $x=L$. Moreover, since their density goes down with increasing $L_x$ (see Fig.~\ref{fig:spectrum}), their effect can be neglected in the thermodynamic limit.

For finite $L_y$, on the other hand, the data shows that the eigenstates are localised as their IPR scales as ${\cal I}_2\sim L_x^0$, except at a certain specific energies.
At these energies, which are of the form $E_n = 4J\cos[n\pi/(L_x+1)]$ with $n=1,2,\cdots,L_x$, the IPR scales as ${\cal I}_2\sim L_x^{-1}$ indicating the delocalised nature of these states
In fact, these states are simply Bloch waves of the form
\eq{
\psi_n(x,y) = \sqrt{\frac{2}{(L_x+1)L_y}}\sin\left(\frac{nx\pi}{L_x+1}\right)\,.
\label{eq:bloch}}
These states can be identified with the dips in the IPR in Fig.~\ref{fig:ipr}(b)

Their origin lies in the fact that the symmetric linear combination of all the orbitals on a given layer couples only to symmetric linear combinations on adjacent layers under the Hamiltonian in Eq.~\eqref{eq:ham}.
Formally, defining $\ket{x}_S = \sum_{y=1}^{L_y}\ket{x,y}/\sqrt{L_y}$, we have
\eq{
H\ket{x}_S = 2J [\ket{x+1}_S+\ket{x-1}_S]\,.
}
As such, these symmetric orbitals are described by a one-dimensional, translation-invariant tight-binding chain which leads to the delocalised Bloch-wave states in Eq.~\eqref{eq:bloch} at the energies mentioned above.
There are $L_x$ such states which constitute a finite fraction, $f_{\rm deloc}=1/L_y$, of the entire spectrum. Moreover since they are delocalised in the background of other localised states, they play a key role in the dynamics as we will discuss shortly.
Since the spectrum has only delocalised Bloch-wave and localised states, both of which independently lead to uncorrelated eigenvalues, the level spacing shows a Poisson distribution.

\paragraph*{Wavepacket dynamics:}
To understand the manifestations of the nature of the spectrum and eigenstates on the dynamics and transport in the model, we study the spreading of an initially localised wavepacket, $\psi(x,y;t=0)= \delta_{x,L_x/2}\delta_{y,L_y/2}$.
Since the notion of locality exists only along $x$, we study the spread of the wavepacket along the $x$ direction via time-evolving distribution
\eq{
\Pi(x,t) =  \sum_{y=1}^{L_y}|\braket{x,y|e^{-iHt}|\psi(t=0)}|^2\,.
}
To quantify the spread, we consider its width
\eq{
\sigma(t) = \left[\sum_{x=1}^{L_x}x^2\Pi(x,t) - \left(\sum_{x=1}^{L_x}x\Pi(x,t)\right)^2\right]^{1/2}\,,
}
which typically grows as $\sigma(t)\sim t^{1/z}$ where $z=1$ and $2$ denote ballistic and diffusive transport, respectively, while $z>2$ corresponds to subdiffusion.

\begin{figure}
\includegraphics[width=\linewidth]{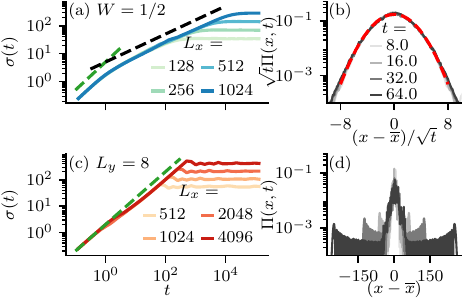} 
\caption{{\bf Wavepacket spreading:} (a) For $W=1/2$, the spreading is diffusive as the width $\sigma(t)\sim t^{1/2}$ (black dashed line) after an initial ballistic transient (green dashed line). (b) The rescaled Gaussian profile of $\Pi(x,t)$ with a width $\sigma(t)$ at different $t$. The red dashed curve shows the corresponding Gaussian profile. (c) For $L_y=8$, the spreading is ballistic (green dashed line) at all timescales. (d) The profile of $\Pi(x,t)$ at different $t$ with a localised core accompanied by the ballistically spreading tails. }
\label{fig:wavepacket}
\end{figure}

The results are shown in Fig.~\ref{fig:wavepacket}. 
For finite $W$, the data in panel (a) show that the wavepacket at the earliest times spreads ballistically due to free dynamics along one-dimensional paths until an $O(1)$ timescale independent of system size. 
At this timescale, the amplitudes from different random paths between any two points on the RLTL graph begin to interfere which leads to the onset of diffusion, marked by $\sigma(t)\sim t^{1/2}$.
This diffusive spreading continues until $\sigma(t)$ saturates to a value $\propto L_x$ due to finite size effects. 
In this diffusive regime, $\Pi(x,t)$ has a conventional Gaussian profile in $x$, $\Pi(x,t) = e^{-(x-\overline{x})^2/2Dt}/\sqrt{2\pi Dt}$ with $\overline{x}=L_x/2$ and $D\approx 5.46$ as shown by the red curve in Fig.~\ref{fig:wavepacket}(b).

On the other hand, the wavepacket dynamics is qualitatively different for finite $L_y$.
As shown in Fig.~\ref{fig:wavepacket}(c), the spreading is ballistic, $\sigma(t)\sim t$, at all timescales until the finite-size saturation.
However, as the profile of $\Pi(x,t)$ in Fig.~\ref{fig:wavepacket}(d) reveals, this transport is anomalous.
Since the initial state overlaps the localised states which have support at $x=L_x/2$, there exists a localised core in $\Pi(x,t)$ at the centre.
Simultaneously, the initial state also overlaps all the Bloch states (in Eq.~\eqref{eq:bloch}) which in fact leads to the ballistic spreading of the tails of the wavepacket. 
Using the properties of translation-invariant, tight-binding chains, it can be straightforwardly shown that $\Pi(x,t)$ has the profile,
\eq{
\Pi(x,t)\!=\!f_{\rm deloc}|{\cal J}_{|x-L_x/2|}(4Jt)|^2 + (1-f_{\rm deloc})g(x)\,,
}
where ${\cal J}_n$ is the Bessel function of the first kind and $g(x)$ is a spatial profile localised around $x=L_x/2$.
The latter obviously does not contribute to the spread and it is the former which leads to a ballistic spread of the wavepacket with a velocity $4J$.

\begin{figure}
\includegraphics[width=\linewidth]{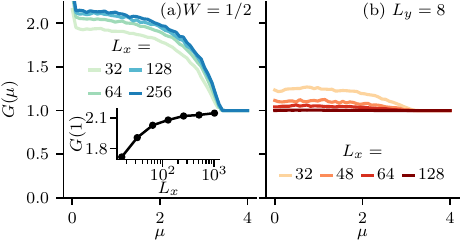} 
\caption{{\bf Conductance}: 
(a) For $W=1/2$ we see a saturation of the conductance (hence conductivity$=G/W$) with increasing $L_x$, implying diffusive transport. The inset shows the saturation with $L_x$ at $\mu=1$. (b) For $L_y=8$, we observe ballistic transport, corresponding to perfect transmission via the Bloch states. For smaller $L_x$, additional contribution to $G(\mu)$ comes from the localised states as their localisation lengths are comparable to $L_x$; this effect vanishes with increasing $L_x$.   
}
\label{fig:conductance}
\end{figure}

\paragraph*{Conductance:} As a further probe of transport we now look at the nonequilibrium current response when the system is connected to leads at different chemical potentials. Specifically we use the non-equilibrium Green's function (NEGF) formalism [see End Matter (EM)] \cite{Datta_1995, ryndyk2016theory, negf}
to compute the zero-temperature linear-response conductance $G(\mu)$ at chemical potential, $\mu$. The asymptotic system size dependence of $G$ on $L_x,L_y$ would characterize the transport behaviour --- in particular, diffusive, ballistic and insulating nature is reflected by the forms $G/L_y \sim  L_x^{-1}, L_x^0$ and $\exp(-c L_x)$, respectively.  In Fig.~\ref{fig:conductance} we show results for the conductance as a function of $\mu$ for the two cases of $W=1/2$ and $L_y=8$. For the former, the inset shows its $L_x$ dependence at $\mu=1$ and the saturation of $G(1)$ implies a finite conductivity($=G/W$), hence confirming diffusive transport. The result for $L_y=8$ shows $L_x$-independent current, which can be shown to be precisely the contribution of the Bloch-wave eigenstates (see EM for details). Note that these Bloch states exist for the $W=1/2$ case as well. Although they constitute a vanishing fraction of the spectrum, they provide a finite contribution to the total conductance since they are ballistic.

\begin{figure}
\includegraphics[width=\linewidth]{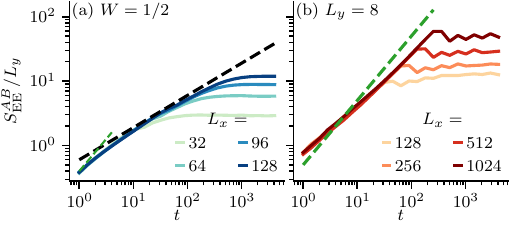} 
\caption{{\bf Entanglement dynamics:} The bipartite EE between two halves of the RLTL graphs each containing half of the layers, for (a) $W=1/2$ and (b) $L_y=8$. In the former, the EE shows a diffusive growth in time, indicated by the black dashed line, whereas in the latter it shows a ballistic growth indicated by the green dashed line.}
\label{fig:entanglement}
\end{figure}

\paragraph*{Entanglement dynamics:}
To complement the results of transport, we also study the bipartite entanglement entropy (EE) between two halves of the graph, the first, $A$, consists of all sites with $x\in [1,L_x/2]$ and the second, $B$, with $x\in[L_x/2+1,L_x]$.
The Von Neumann EE between $A$ and $B$ is given by~\cite{peschel2009reduced} 
\eq{
S_{\rm EE}^{AB} = -\sum_{k}[\lambda_k\ln\lambda_k +(1-\lambda_k)\ln(1-\lambda_k)]\,,
}
where $\{\lambda_k\}$ is the set of eigenvalues of the reduced correlation matrix $[C_A]_{(x,y);(x',y')}=\braket{c_{x,y}^\dagger c_{x',y'}}$ with $(x,y)\in A$.
We start with an initial state which is a product state with $N/2$ sites randomly filled across the graph. 
The results for $S_{\rm EE}^{AB}$ are shown in Fig.~\ref{fig:entanglement} where it is evident that their behaviour closely mirrors that of wavepacket spreading. 
For $W=1/2$, $S_{\rm EE}^{AB}$ grows diffusively in time after a very early time ballistic transient whereas for a fixed $L_y=8$, $S_{\rm EE}^{AB}$ grows ballistically. 
Also note that for $W=1/2$, in the universal growth, the data for different $L_x$ collapse onto each other when rescaled by $L_y$.
This is simply because the interface between the two halves of the graph is of length $L_y$ and hence the EE at finite time is proportional to it.

To understand the results, it is useful to exploit the Gaussian nature of the states which lets us connect $S_{\rm EE}^{AB}$ to the cumulants of particle number in the subsystem~\cite{klich2009quantum}
\eq{
S_{\rm EE}^{AB} = \sum_{k=1}^\infty\frac{(2\pi)^{2k}|B_{2k}|}{(2k)!}\chi_{2k}\,,
\label{eq:SEE-series}
}
where $\chi_m$ and $B_m$ are the $m^{\rm th}$ cumulant and Bernoulli number.
For $L_x\gg 1$, the sum in Eq.~\eqref{eq:SEE-series} is dominated by the first term $S_{\rm EE}^{AB}\approx \pi^2\chi_2/3$ where $\chi_2\equiv (\Delta N_A)^2$ is the fluctuation in the particle number in $A$. 
This can be exactly related to the wavepacket spreading as follows. Defining $\Pi(\mathbf{r},t; \mathbf{r_0})$ as the probability of finding a single particle at $\mathbf{r}$ at time $t$ initialised at $\mathbf{r_0}$, $Q_A(t;\mathbf{r_0})\equiv \sum_{\mathbf{r}\in A}\Pi(\mathbf{r},t; \mathbf{r_0})$ is the total probability of finding the particle in subsystem $A$. 

Averaging over all initial half-filled, product states, we obtain (see EM for details)
\eq{
\overline{\braket{(\Delta N_A(t))^2}} = \frac{1}{4}\sum_{\mathbf{r_0}}Q_A(t;\mathbf{r_0})[1-Q_A(t;\mathbf{r_0})]\,,
\label{eq:fluc-wp}
}
These relations show that the EE is well described by the number of particles sloshing back and forth across the cut. 
If the wavepacket spreads as $\sim t^{1/z}$, then particles that are within a distance $\sim t^{1/z}$ from the cut have a finite probability of crossing the cut and contributing to the entanglement. 
Since the length of the cut is $L_y$, the EE therefore scales as $\sim L_y t^{1/z}$ and therefore mirrors the wavepacket spreading. 

\paragraph*{Outlook:}
In summary, we have demonstrated that geometric randomness can lead to quantum chaos and diffusive transport in non-interacting systems with otherwise uniform couplings. 
The interplay of this with disordered onsite potentials or disordered hopping amplitudes is naturally a question for the future. 
Since RLTL graphs share features of both finite dimensional graphs where the Anderson transition is known to be a continuous transition~\cite{evers2008anderson}, and high-dimensional tree-like graphs where the transition is Kosterlitz-Thouless-like~\cite{garciamata2017scaling,garciamata2022critical}, the question of the existence, and its nature, of the Anderson transition on RLTL graphs is an extremely interesting one.
Also, since the disorder-free RLTL graphs in the quasi one-dimensional limit host both localised and delocalised states, the fate of the model upon subjecting it to periodic driving is an interesting question. In particular, it will be interesting to see if the driving-induced hybridisation between the localised states mediated by the delocalised states leads to robust multifractality analogous to periodically-driven quasiperiodic systems with mobility edges~\cite{roy2018multifractality}.

\begin{acknowledgments}
{\it Acknowledgements:} The authors would like to thank K. Damle,  D. Dhar and J. Radhakrishnan for many valuable discussions.
This work was supported by the Department of Atomic Energy, Government of India, under project nos. RTI4019 and RTI4013. A.D. acknowledges the J.C. Bose
Fellowship (JCB/2022/000014) of the Science and Engineering Research Board of the Department of Science and Technology (SERB-DST), Government of India.
S.R. acknowledges support from SERB-DST (India) under Grant No. SRG/2023/000858, from ANRF (India) under Grant No. ANRF/ARG/2025/004045/PS, and from a Max Planck Partner Group grant between ICTS-TIFR, Bengaluru and MPIPKS, Dresden. 

\paragraph*{Note:} During the completion of this work, we became aware of Ref.~\cite{sarkar2026emergentquantumchaoscorrelations} which also addresses geometrically random graphs. Our results, where overlapping, are consistent with them.
\end{acknowledgments}

\bibliography{references}

\clearpage

\onecolumngrid

\section{End Matter}

\subsection{Details of NEGF calculations}
Here we discuss the details of our conductance calculations using the NEGF formalism. In our setup, we attach a semi-infinite 1d lead to each of the sites at the boundary layers ($x=1$ and $L_x$) of the RLTL graph and compute the conductance $G(\mu)$ between the left and right sets of leads at a chemical potential $\mu$. We model the leads attached at the boundary layer site $y$ via  nearest-neighbour tight-binding (NNTB) Hamiltonians $H_L^{(y)}$ and $H_R^{(y)}$ on 1d chains that couple to the graph via the coupling matrices $V_{SL}$ and $ V_{SR}$ described by 
\begin{align}
    &H_L^{(y)}=  2J\sum_{x=-\infty}^{-1} 
    b_{x,y}^
    \dagger b_{x+1,y}+ {\rm h.c}\ , \ 
   \hspace{0.3 cm}H_R^{(y)}=  2J\sum_{x=L_x+1}^{\infty} 
    b_{x,y}^
    \dagger b_{x+1,y}+ {\rm h.c} \ ,    \hspace{0.3cm} V_{SL}= V_{SR}= 2J\sum_{y=1}^{L_y}  c_{\nu,y}^\dagger b_{\sigma,y} +{\rm h.c}
    \label{eq: lead-H}
\end{align}
where $b_{x,y}$ is the annihilation operator acting on the leads at site $y$ of the layer $x$. Also, $\nu=1,L_x$ and $\sigma =0, L_x+1$ for the left and right leads, respectively. The full lead Hamiltonians are $H_\chi =\sum_{y=1}^{L_y}H_\chi^{(y)}$ with $\chi=L,R$.  Note that we take the hopping amplitudes within the leads and that for the system-lead coupling to be twice that of those within the system. 
Courtesy this choice,
the Bloch-wave modes transmit perfectly 
across the system, {\it i.e.} $G_{\rm Bloch} (\mu)=1$ for $\mu <|4J|$ and therefore the total conductance $G(\mu)= G_{\rm Bloch}(\mu)+ G_{\rm random} (\mu)$, where $G_{\rm random} (\mu)$ is the contribution to the total conductance due to the remaining random states within the system.

 The leads support the propagating modes within $E_q=4J \cos q$ with $q \in [0, \pi]$ with $E=\pm 4J$ the extreme energies of their spectrum. Thus, the conductance $G(\mu)$ vanishes for $|\mu| > 4J$.  Moreover, it is worth noting that the leads are also bipartite in the same sense as the RLTL graph, resulting in symmetric conductance $G(\mu)$ about $\mu =0$. Hence, it is sufficient to look at $G(\mu)$ for $\mu \in [0, 4J]$ and in the subsequent discussion, we restrict ourselves to this range (Fig. \ref{fig:conductance}).
Within this formalism, the contribution of the leads enters through self-energy matrices $\Sigma_\chi^{\pm}(\mu)= V_{S\chi} \mathcal{G}_{\chi s}^\pm (\mu) V_{S\chi}^\dagger $ 
and 
\begin{align}
    [\mathcal{G}_{
    \chi s}^\pm (\mu)]_{y'y''}=\langle 
    \sigma,  y' |\left[(\mu\pm i0^+)\mathbb{I}- H_\chi\right]^{-1}|
    \sigma, y'' \rangle= \dfrac{\delta_{y'y''}}{2J}\left[ \dfrac{\mu}{2}\mp i \sqrt{1-\dfrac{\mu^2}{16J^2}}\right]  
\end{align}
represents the matrix element of the $L_y \times L_y$ boundary Green's function between the sites $y'$ and $y''$ of the lead  $\chi$ on the boundary at $\sigma$.  
We compute
the zero-temperature linear response conductance $G(\mu)$ across the system at a chemical potential $\mu$ using the standard NEGF expression
\begin{align}
    G(\mu)= 
    \operatorname{Tr} [\Gamma_L (\mu)\mathcal{G}_{\rm eff}^+ (\mu) \Gamma_R  (\mu )\mathcal{G}_{\rm eff}^- (\mu)]
\end{align}
where 
\begin{align}
    \mathcal{G}_{\rm eff}^{\pm}(\mu)=\left[\mu \mathbb{I}-H-\Sigma^\pm_{L}(\mu)-\Sigma^\pm_R(\mu)\right]^{-1} ; \hspace{2cm}  \Gamma_\chi (\mu)= i\left[\Sigma^+_{\chi}(\mu)-\Sigma^-_{\chi}(\mu)\right].
\end{align}

It is important to note that the conductance for the quasi one-dimensional limit of the RLTL graph is solely governed by the robust Bloch wave modes (defined in Eq. \eqref{eq:bloch}), leading to a ballistic transport. This is because all other states in this case are localised and, therefore, do not contribute to the transport. However, since in our setup these Bloch wave modes transmit perfectly, the conductance in this limit would thus converge to $G(\mu) \to 1$ in the thermodynamic limit ($L_x\to \infty$).

\subsection{Number fluctuations and wavepacket spreading}
In this section of the EM, we derive the relation between particle number fluctuations in the subsystem and wavepacket spreading mentioned in Eq.~\eqref{eq:fluc-wp}.
We will consider the subsystem $A$ to contain all the sites on layers $x=1$ through $L_A$. 
To start, we note that the number fluctuation in subsystem $A$ can be expressed as $\braket{(\Delta N_A)^2} = {\rm Tr}[C_A] - {\rm Tr}[C_A^2]$ which using the form of two-point fermionic correlations can be written as 
\eq{
\braket{(\Delta N_A)^2} = \sum_{\mbf{r}\in I_0}Q_A(t,\mbf{r})[1-Q_A(t,\mbf{r})] - \sum_{\substack{{\mbf{r}\neq\mbf{r'}}\\\mbf{r},\mbf{r'}\in I_0}}\left|R_A(t,\mbf{r},\mbf{r'})\right|^2\,,
\label{eq:NA-sq-exp}
}
where $Q_A(t,\mbf{r}) = \sum_{\mbf{r}_A\in A}|G_{\mbf{r}\mbf{r}_A}(t)|^2$  and $R_A(t,\mbf{r},\mbf{r'}) = \sum_{\mbf{r}_A\in A}G_{\mbf{r}\mbf{r}_A}(t)G^\ast_{\mbf{r'}\mbf{r}_A}(t)$ with the single-particle propagator $G_{\mbf{r}\mbf{r'}}(t) = \braket{\mbf{r'}|e^{-iHt}|\mbf{r}}$ and $I_0$ denotes the set of all sites that were initially occupied.
We next average over all such sets of initial conditions, which is equivalent to averaging over all the choices of $I_0$.
To this end, we note that 
\eq{
\overline{\sum_{\mbf{r}\in I_0}[\cdots]} = \frac{1}{2}\sum_{\mbf{r}}[\cdots]\,,~
\overline{\sum_{\substack{{\mbf{r}\neq\mbf{r'}}\\\mbf{r},\mbf{r'}\in I_0}}[\cdots]}=\frac{L_xL_y-2}{4(L_xL_y-1)}\sum_{\mbf{r}\neq\mbf{r'}}[\cdots]\,,
\label{eq:avg-I0}
}
where the overline denotes the averaging over $I_0$.
Using this, the average of the first term on the right-hand side of Eq.~\eqref{eq:NA-sq-exp} can be expressed as 
\eq{
\overline{\sum_{\mbf{r}\in I_0}Q_A(t,\mbf{r})[1-Q_A(t,\mbf{r})]} = \frac{1}{2}\sum_{\mbf{r}}Q_A(t,\mbf{r})[1-Q_A(t,\mbf{r})]\,.
\label{eq:avg-1}
}
It will also be useful shortly to note that
$\sum_{\mbf{r}}Q_A(t,\mbf{r}) = L_AL_y$. Averaging the second term on the right-hand side of Eq.~\eqref{eq:NA-sq-exp} using Eq.~\eqref{eq:avg-I0}, we obtain
\eq{
\overline{\sum_{\substack{{\mbf{r}\neq\mbf{r'}}\\\mbf{r},\mbf{r'}\in I_0}}\left|R_A(t,\mbf{r},\mbf{r'})\right|^2} &= \frac{L_xL_y-2}{4(L_xL_y-1)}\left[
\sum_{\mbf{r}_A,\mbf{r}_A^\prime\in A}\left|\sum_\mbf{r}G_{\mbf{r}\mbf{r}_A}(t)G_{\mbf{r}\mbf{r}_A^\prime}^\ast(t)\right|^2 - \sum_{\mbf{r}}\left|\sum_{\mbf{r}_A\in A}|G_{\mbf{r}\mbf{r}_A}(t)|^2\right|^2\right]\\
&=\frac{L_xL_y-2}{4(L_xL_y-1)}\left[L_AL_y - \sum_{\mbf{r}}Q_A^2(t,\mbf{r})\right]\overset{L_x\gg 1}{\approx}\frac{1}{4}\sum_{\mbf{r}}Q_A(t,\mbf{r})[1-Q_A(t,\mbf{r})]\,,\label{eq:avg-2}
}
where in the second line we used the unitarity condition $\sum_{\mbf{r''}}G_{\mbf{r}\mbf{r''}}(t)G^\ast_{\mbf{r'}\mbf{r''}}(t) = \delta_{\mbf{r}\mbf{r'}}$ and also identified $L_AL_y=\sum_{\mbf{r}}Q_A(t,\mbf{r})$.
Using the results in Eqs.~\eqref{eq:avg-1} and \eqref{eq:avg-2} back into Eq.~\eqref{eq:NA-sq-exp} we have 
\eq{
\overline{\braket{(\Delta N_A)^2}} = \frac{1}{4}\sum_{\mbf{r}}Q_A(t,\mbf{r})[1-Q_A(t,\mbf{r})]\,,
}
which is precisely the relation in Eq.~\eqref{eq:fluc-wp}.

\end{document}